\documentclass[twocolumn,showpacs,preprintnumbers,amsmath,amssymb]{revtex4-1}

\usepackage{graphicx}
\usepackage{dcolumn}
\usepackage{bm}

\begin{document}

\preprint{}

\title{
Two-loop level rainbowlike supersymmetric contribution to the fermion electric dipole moment 
}

\author{Nodoka Yamanaka}
\affiliation{%
Research Center for Nuclear Physics, Osaka University, Ibaraki, Osaka 567-0047, Japan}%

\date{\today}

\begin{abstract}
We calculate the two-loop level electric and chromo-electric dipole moments of the fermion involving the fermion-sfermion inner loop, gaugino, and higgsino in the minimal supersymmetric standard model, and analyze the chromo-electric dipole moment with the top-stop inner loop.
It is found that this contribution is comparable with, and even dominates, in some situations over the Barr-Zee type diagram generated from the $CP$ violation of the top squark sector in TeV scale supersymmetry breaking.
\end{abstract}

\pacs{12.60.Jv, 11.30.Er, 13.40.Em, 14.80.Ly}
\maketitle


The search for supersymmetry \cite{mssm} as the leading candidate of new physics beyond the standard model is currently performed within many experimental approaches, and many active studies have been done so far.
One promising way to search for supersymmetry is the measurement of the electric dipole moment (EDM).
The standard model prediction of the EDM is very small, so that any observations of finite EDM give direct evidence of the $CP$ violation of new physics \cite{smedm,pospelovreview}.

The minimal supersymmetric standard model (MSSM) has many $CP$ phases, and many analyses with natural $CP$ violating parameters predict large EDM at the one-loop level \cite{susyedm1-loop,pospelovreview}, but these scenarios confront a serious difficulty in explaining the strong experimental limits of the EDM of many systems such as the neutron \cite{baker} or the electron \cite{hudson}.
These stringent constraints lead us to think of supersymmetric models with no $CP$ violation in the first and second generation of sfermions, and the source of the $CP$ nonconservation is left only to the third generation sector.

\begin{figure}[htb]
\includegraphics[width=8cm]{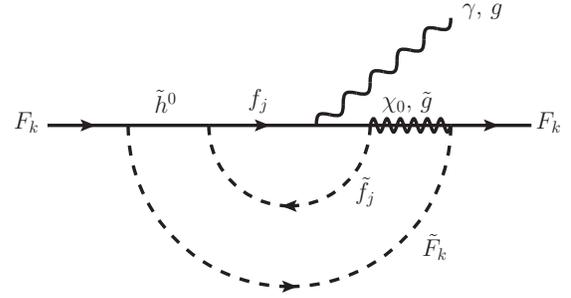}
\caption{\label{fig:rainbow} 
Example of rainbow diagram contributing to the fermion EDM in MSSM.
The neutralino is given by $\chi_0$ and the gluino by $\tilde g$.
}
\end{figure}

The $CP$ violation of the third generation sfermions contributes to the fermion EDM through the two-loop level process.
Previous studies of the Barr-Zee type diagrams showed that this contribution is actually sensitive to the $CP$ violation of the third generation in supersymmetric models \cite{barr-zee}.
In this work, we extend the two-loop level analysis of the supersymmetric effects to the fermion EDM. 
For that, we calculate the two-loop level {\it rainbowlike} diagrams which involve both the higgsino and gaugino, as shown in Fig. \ref{fig:rainbow}.
This contribution has a fermion-sfermion inner loop, connected to the external fermion line with a higgsino and a gaugino, and the flavor structure is exactly the same as for the Barr-Zee type diagram, with the same coupling constants.
The potential importance of these rainbow diagrams was pointed in Ref. \cite{pilaftsis}, arguing that they can actually have comparable size with the supersymmetric Barr-Zee type contribution.
We should therefore analyze this process in detail.

In this paper, we first calculate the rainbow diagram contribution to the fermion EDM in the MSSM and give its explicit formula.
We then analyze the effect of the chromo-EDM of the rainbow diagrams and compare with that of the Barr-Zee type diagrams.


Let us first give the lagrangian of the particles relevant in this discussion.
In this work, we assume that the supersymmetry breaking scale is beyond TeV, which is strongly suggested by the recent result of the LHC experiment \cite{lhc}, and we neglect the effects proportional to the Higgs vacuum expectation value against the sparticle masses.
This assumption especially simplifies the calculation of the neutralino contribution.
The higgsino and the gaugino sectors of the neutralino mass matrix are completely separated, since the mixing with each other occurs only through the gauge interaction with the Higgs vacuum condensation.
The mass matrix of the neutralino is then given by
\begin{equation}
{\cal L}_{}\chi_0 = -\frac{1}{2} \bar \chi_ {0R} M_N \chi_{0L} + {\rm H.c.}\, ,
\end{equation}
with
\begin{equation}
M_N
\approx
\left(
\begin{array}{cccc}
0&|\mu | e^{i\theta_\mu} &0&0\cr
| \mu | e^{i\theta_\mu} &0 &0&0\cr
0&0 &|m_{\lambda_1} | e^{i\theta_1}&0\cr
0&0 &0&|m_{\lambda_2}| e^{i\theta_2}\cr
\end{array}
\right)\, ,
\label{eq:neutralinomassmatrix}
\end{equation}
where the corresponding neutralino field vector $\chi_0$ has as its first two components the up- and down-type higgsinos, and the last two components refer to the $U(1)_Y$ and the neutral $SU(2)_L$ gauginos.
The components of the mass matrix have $CP$ phases in the general case [see the complex phases $\theta_\mu$, $\theta_1$ and $\theta_2$, Eq. (\ref{eq:neutralinomassmatrix}) ], and these can manifest themselves as an observable effect when mass insertions occur in the process.
For the higgsino however, the up- and down-type fields are fully mixed during the propagation with two mass eigenstates taking the same absolute mass eigenvalue, but with {\it opposite sign}.
This fact exhibits a complete cancellation of the mass insertion for the higgsino, and the $CP$ phase $\theta_\mu$ is irrelevant in the higgsino propagation.
Moreover, the higgsino cannot mix the up- and down-type components due to the equal (absolute) mass eigenvalues.
Thus, the EDM of a fermion with isospin $+(-)\frac{1}{2}$ can have only an inner loop contribution with an isospin $+(-)\frac{1}{2}$ fermion (for example, the charged lepton cannot have a top-stop inner loop).
For the gluino, we remove its $CP$ phase as usual.

Let us mention shortly the situation where the Higgs vacuum expectation value is not neglected.
In this case, the mixing between gaugino and higgsino during the neutralino propagation occurs.
This modifies the chiral structure of the inner loop or the external fermion line, since one gaugino-matter vertex is replaced by a higgsino-matter vertex (or vice versa).
To keep the chiral structure of the EDM, we must add an additional insertion of interaction with Higgs vacuum condensation [higgsino-gaugino transition, mass insertion of matter fermion, or gauge eigenstate transition of sfermion ($\tilde f_L\leftrightarrow \tilde f_R$ transition)] somewhere in the same process.
The effect of the Higgs vacuum expectation value is therefore a correction of order $O(e^2 v^2 / \Lambda_{\rm susy}^2) $, with $e$ the electromagnetic coupling constant, $v$ the Higgs vacuum expectation value, and $\Lambda_{\rm susy}$ the supersymmetry breaking scale.
The Higgs vacuum condensation also mixes the up- and down-type higgsinos, due to their transition into the gaugino in the intermediate state.
This mixing is also a correction of $O(e^2 v^2 / \Lambda_{\rm susy}^2) $ with an additional suppression due to the  factor $\cos \beta$ (for large $\tan \beta$).

The sfermion mass matrix is given by
\begin{equation}
M_{\tilde f}^2
\approx
\left(
\begin{array}{cc}
m_{\tilde f_L}^2 & m_f (A_f^* - \mu R_f) \\
m_f (A_f - \mu^* R_f) & m_{\tilde f_R}^2 \\
\end{array}
\right)\ ,
\end{equation}
where $R_f = \cot \beta$ for up-type quarks, and $R_f = \tan \beta$ for down-type quarks and charged leptons.
We neglect the off-diagonal components of the first and second generation sfermion mass matrices.
In the following, we will use the unitary matrix to obtain the mass eigenbasis for the third generation sfermion.
Explicitly, the unitary matrix with angles $\theta_f$ and $\delta_f$ is given by the following relation \cite{pilaftsis2}
\begin{equation}
\left(
\begin{array}{c}
\tilde f_L \\
\tilde f_R \\
\end{array}
\right)
= 
\left(
\begin{array}{cc}
1&0 \\
0& e^{i\delta_f} \\
\end{array}
\right)
\left(
\begin{array}{cc}
\cos \theta_f & \sin \theta_f \\
-\sin \theta_f & \cos \theta_f \\
\end{array}
\right)
\left(
\begin{array}{c}
\tilde f_1 \\
\tilde f_2 \\
\end{array}
\right) \ .
\end{equation}
The CP phase is given by $\delta_f = \arg (A_f - R_f \mu^*)$.

We should also give the gaugino-sfermion-fermion and higgsino-sfermion-fermion interactions.
These are, respectively, given by
\begin{eqnarray}
{\cal L}_{\lambda}
&=&
\sum_{\tilde f} \sqrt{2} g^{(n)}_{\tilde f_{L/R}} \tilde f_{L/R}^\dagger t_a \bar \lambda_{n,a} P_{L/R} f +{\rm H.c.}
\ ,
\nonumber\\
{\cal L}_{\tilde h} &=&
-\sum_f Y_f \left[\, 
\overline{ \tilde h_i^{0c}} P_L f \tilde f_R^\dagger
+\bar f P_L\tilde h_i^0 \tilde f_L
 \, \right] + {\rm H.c.} 
\, ,
\end{eqnarray}
where $\tilde h_i^0$ is the up- (down-)type higgsino for the up-type quark (down-type quark or charged lepton).
The index $a$ is the gauge index, but will be omitted from now.
The matter-Higgs Yukawa couplings $Y_f$ satisfy $Y_f = \frac{m_f e}{\sqrt{2} \sin \theta_W m_W r_f} $ where $r_f=\sin \beta\ (\cos \beta) $ for up-type quarks (down-type quarks and charged leptons).
The convention for the sign of the gauge coupling is $D^\mu \equiv \partial^\mu -ig A_a^\mu t_a$ where $t_a$ is the generator of the gauge group.
The index $n$ gives the gauge group of the gaugino $\lambda_n$.
We define $\lambda_1$ as the $U(1)_Y$ gaugino, $\lambda_2$ as the neutral $SU(2)_L$ gaugino, and $\lambda_3 = \tilde g$ as the gluino.
The fermion-sfermion-gaugino coupling constants are given as follows:
for the gluino couplings, we have
$g^{(3)}_{\tilde q_L} = g_s$ and
$g^{(3)}_{\tilde q_R} = -g_s$.
For the couplings of $\lambda_1$, we have
$g^{(1)}_{\tilde q_L} = \frac{1}{6}\frac{e}{\cos \theta_W}$, 
$g^{(1)}_{\tilde u_R} = -\frac{2}{3}\frac{e}{\cos \theta_W}$,
$g^{(1)}_{\tilde d_R} = \frac{1}{3}\frac{e}{\cos \theta_W}$,
$g^{(1)}_{\tilde l_L} = -\frac{1}{2}\frac{e}{\cos \theta_W}$, and
$g^{(1)}_{\tilde e_R} = \frac{e}{\cos \theta_W}$.
We have finally for the $\lambda_2$ couplings 
$g^{(2)}_{\tilde u_L} = \frac{1}{2}\frac{e}{\sin \theta_W}$,
$g^{(2)}_{\tilde d_L} = -\frac{1}{2}\frac{e}{\sin \theta_W}$, and
$g^{(2)}_{\tilde l_L} = -\frac{1}{2}\frac{e}{\sin \theta_W}$.


\begin{figure}[htb]
\begin{center}
\includegraphics[width=8cm]{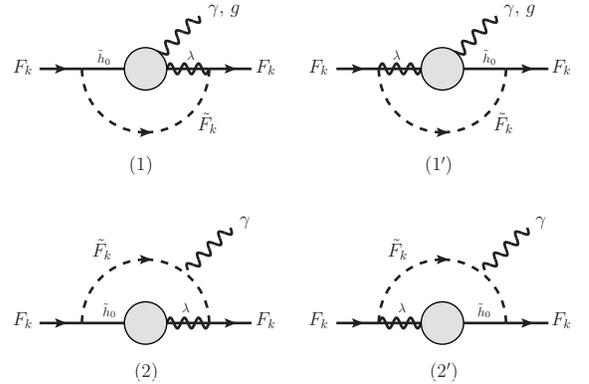}
\caption{\label{fig:sec_loop}
Possible insertions of one-loop effective vertex for the rainbow diagram in MSSM.
The grey blob represents the one-loop effective vertex.
}
\end{center}
\end{figure}

We are now ready to calculate the rainbow diagrams.
There are two possible insertions of a one-loop level effective vertex to form rainbow diagrams (see Fig. \ref{fig:sec_loop}).

\begin{figure}[htb]
\includegraphics[width=7.5cm]{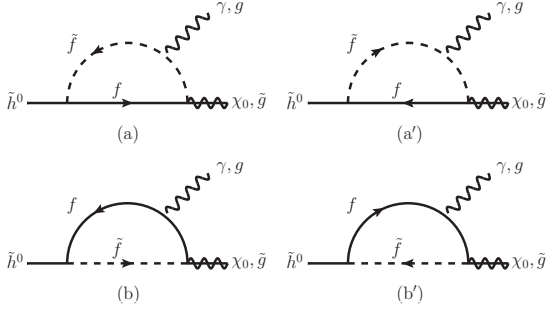}
\caption{\label{fig:hgg} 
One-loop contribution to the effective higgsino-gaugino-gauge boson vertex.
}
\end{figure}

The first contribution is the insertion of the one-loop effective higgsino-gaugino-gauge boson vertex [see Fig. \ref{fig:sec_loop} (1) and (1')].
The inner one-loop diagrams are depicted in Fig. \ref{fig:hgg}.
This contribution can generate both EDM and chromo-EDM.

\begin{figure}[htb]
\begin{center}
\includegraphics[width=7.5cm]{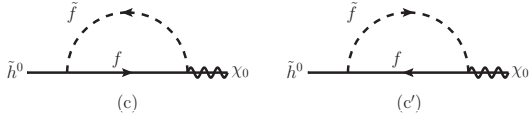}
\caption{\label{fig:hg}
One-loop diagram contribution to the higgsino-gaugino transition.
}
\end{center}
\end{figure}

The second type of contribution is the insertion of the one-loop effective higgsino-gaugino transition [see Fig. \ref{fig:sec_loop} (2) and (2')].
The contributing one-loop diagrams are shown in Fig. \ref{fig:hg}.
Note that this type of rainbow diagram does not contribute to the chromo-EDM due to the gauge invariance.

To obtain the EDM contribution, the above rainbow diagrams are calculated to the first order in the external momentum carried by the gauge boson (this manipulation is suitable since the EDM is the first order coefficient of the multipolar expansion).
In our calculation, we have neglected the fermion mass originating from the mass insertion of the inner loop.
We obtain then the following formula for the EDM of the light fermion $F$:
\begin{equation}
d_F = d^1_F + d^2_F \ ,
\end{equation}
with 
\begin{eqnarray}
d_F^1
&\approx &
\sum_f
\frac{n_c Q_f Y_f Y_F}{64\pi^3} \sum_{n=1,2} |m_{\lambda_n}| \sin (\delta_f -\theta_n ) 
\nonumber\\
&&
\times \frac{e (g^{(n)}_{\tilde f_L} -g^{(n)}_{\tilde f_R})  }{4\pi} \sin \theta_f \cos \theta_f  
\nonumber\\
&& 
\times \sum_{\tilde F= \tilde F_L,\tilde F_R} \hspace{-1em} s \, g^{(n)}_{\tilde F}
\Bigl[ 
F' (|m_{\lambda_n}|^2 , |\mu|^2 , m_{\tilde F}^2 , m_{\tilde f_{2}}^2)
\nonumber\\
&& \hspace{7em}
-F' (|m_{\lambda_n}|^2 , |\mu|^2 , m_{\tilde F}^2 , m_{\tilde f_{1}}^2)
\Bigr]
,
\label{eq:edmhgg}
\\
d_F^2
&=&
\sum_f
\frac{n_c Q_F Y_f Y_F }{64\pi^3}  \sum_{n=1,2} |m_{\lambda_n}| \sin (\delta_f -\theta_n) 
\nonumber\\
&&
\times \frac{ e (g^{(n)}_{\tilde f_L} +g^{(n)}_{\tilde f_R})}{4\pi} \sin \theta_f \cos \theta_f 
\nonumber\\
&& 
\times \sum_{\tilde F= \tilde F_L,\tilde F_R} \hspace{-1em}  g^{(n)}_{\tilde F} m_{\tilde F}^2 
\Bigl[ 
F''(|m_{\lambda_n}|^2 , |\mu|^2 , m_{\tilde F}^2 , m_{\tilde f_{1}}^2 )
\nonumber\\
&&
\hspace{7.5em}
-F''(|m_{\lambda_n}|^2 , |\mu|^2 , m_{\tilde F}^2 , m_{\tilde f_{2}}^2 )
\Bigr]
 ,\ \ 
\label{eq:edmhg}
\end{eqnarray}
where $n_c=3$ for the inner quark-squark loop and $n_c=1$ for the inner lepton-slepton loop. 
Here the EDM $d^1$ is the contribution from the first type of insertion (with one-loop effective higgsino-gaugino-gauge boson vertex), and $d^2$ the EDM generated by the second type insertion (with one-loop effective higgsino-gaugino transition).
The fields $\tilde f_1$ and $\tilde f_2$ are the mass eigenstates of the sfermion $\tilde f$.
The constant $s$ is $+1$ for left-handed sfermion $\tilde F_L$ and $-1$ for right-handed sfermion $\tilde F_R$.
The electric charge of the inner loop fermion (sfermion) and the external fermion in the unit of $e$ is denoted, respectively, by $Q_f$ and $Q_F$.
The functions $F'$ and $F''$ are defined by
\begin{widetext}
\begin{eqnarray}
F' (a,b,c,f)
&\equiv &
\frac{a \ln a - f \, {\rm Li}_2 \left( 1 - \frac{a}{f} \right)}{(a-b)(a-c)}
+\frac{b \ln b - f \, {\rm Li}_2 \left( 1 - \frac{b}{f} \right)}{(b-a)(b-c)}
+\frac{c \ln c - f \, {\rm Li}_2 \left( 1 - \frac{c}{f} \right)}{(c-a)(c-b)}
\, ,
\\
F''(a,b,c,f)
&\equiv &
 a^2 R_{13}(a,b,c) J_1(a,f)
+b^2 R_{13}(b,a,c) J_1(b,f)
+(a^2 b^2 - 3abc^2 +ac^3 +bc^3 ) R_{33} (c,a,b) J_1 (c,f)
\nonumber\\
&&
+c(ac+bc-2ab) R_{22} (c,a,b) J_2 (c,f) 
+c^2 R_{11} (c,a,b ) J_3 (c,f)
\nonumber\\
&&-f (1+2\ln f) 
\Bigl[
 a\ln a R_{13} (a,b,c)
+b\ln b R_{13} (b,a,c)
+(a^2 b +ab^2 -3abc +c^3) \ln c R_{33} (c,a,b)
\nonumber\\
&& \hspace{6em} 
+\frac{1}{c}(ab-c^2) R_{22} (c,a,b) 
-\frac{1}{2c} R_{11} (c,a,b)
\Bigr]
\nonumber\\
&&
-2f \left[
 a R_{13} (a,b,c) J_1 (a,f)
+b R_{13} (b,a,c) J_1 (b,f)
+(a^2b+ab^2 -3abc +c^3 ) R_{33} (c,a,b) J_1 (c,f) 
\right.
\nonumber\\
&& \hspace{4em}
\left.
+(c^2-ab) R_{22}(c,a,b) J_2 (c,f)
+c R_{11}(c,a,b) J_3 (c,f)
\right]
\nonumber\\
&&+f^2 \ln f 
\Bigl[
 \ln a R_{13} (a,b,c)
+\ln b R_{13} (b,a,c)
+\ln c (a^2 +b^2 +ab -3ac -3bc +3c^2 ) R_{33} (c,a,b)
\nonumber\\
&&
\hspace{5em} 
+\frac{1}{c} (a+b-2c) R_{22} (c,a,b)
-\frac{1}{2c^2} R_{11} (c,a,b)
\Bigr]
\nonumber\\
&&
+f^2 \left[
 R_{13} (a,b,c) J_1 (a,f)
+R_{13} (b,a,c) J_1 (b,f)
+(a^2 +b^2 +ab -3ac -3bc +3c^2 ) R_{33} (c,a,b) J_1 (c,f)
\right.
\nonumber\\
&&\hspace{3em}
\left.
-(a+b-2c) R_{22} (c,a,b) J_2 (c,f)
+R_{11} (c,a,b) J_3 (c,f)
\right] \ ,
\end{eqnarray}
\end{widetext}
with ${\rm Li}_2 (x)$ denoting the dilogarithm function, $R_{nm} (a,b,c) \equiv \frac{1}{(a-b)^n (a-c)^m}$, $J_1 (a,f) \equiv {\rm Li}_2 (1-a/f ) +\frac{1}{2} \ln^2 f -\ln a \ln f$, $J_2 (c,f) \equiv \frac{c \ln c -f \ln f}{c(c-f)}$ and $J_3 (c,f) \equiv \frac{c^2 \ln c - c(c-f) -f (2c-f) \ln f}{2c^2 (c-f)^2}$.

We also obtain the following expression for the quark chromo-EDM:
\begin{eqnarray}
d_F^c
&\approx &
\sum_f
\frac{Y_f Y_F}{64\pi^3} \alpha_s g_s \sin \theta_f \cos \theta_f  
\sin \delta_f m_{\tilde g} 
\nonumber\\
&& \times
\sum_{\tilde F=\tilde F_L ,\tilde F_R}
\Bigl[ 
F' (m_{\tilde g}^2 , |\mu|^2 , m_{\tilde F}^2 , m_{\tilde f_{2}}^2)
\nonumber\\
&& \hspace{6em}
-F' (m_{\tilde g}^2 , |\mu|^2 , m_{\tilde F}^2 , m_{\tilde f_{1}}^2)
\Bigr]
.
\label{eq:cedm}
\end{eqnarray}
We see from Eqs. (\ref{eq:edmhgg}), (\ref{eq:edmhg}), and (\ref{eq:cedm}) that the rainbow diagram contribution vanishes if the mass eigenvalues of the inner loop sfermion $m_{\tilde f_{1}}$ and $m_{\tilde f_{2}}$ are degenerate.
We have also encountered this property in the analysis of the supersymmetric Barr-Zee type diagram \cite{barr-zee}.

Let us now move to the analysis.
Here we analyze the effect of the chromo-EDM of quarks since it gives the leading $CP$-odd dipole moment at the two-loop level.
We explicitly choose the following sparticle masses and parameters, with an overall factor of supersymmetry breaking scale $\Lambda_{\rm susy}$:
$m_{\tilde g} = 1.8\times \Lambda_{\rm susy}$, $m_{\tilde t_{1}} = 1.5\times \Lambda_{\rm susy}$, $m_{\tilde t_{2}} = 1.6\times \Lambda_{\rm susy}$.
The up squark masses for both gauge eigenstates are given by $m_{\tilde u_L} =m_{\tilde u_R} = 1.7\times \Lambda_{\rm susy}$.
For the $\mu$ parameter, we do not know the origin, but it must be relatively heavy, as it gives the mass of the higgsino.
In our discussion, we fix $|\mu| = 1$ TeV.
We have also fixed $\tan \beta =40$.
To compare with the Barr-Zee type contribution, we need also the mass of the $CP$-odd Higgs boson.
The mass of the Higgs boson at the tree level is given by $M_a^2 = b (\tan \beta + \cot \beta) $, where $b$ is the soft breaking parameter with mass dimension 2.
We assume $b=(1\times \Lambda_{\rm susy} )^2$.
We must note that under this assumption, the large enhancement of the Barr-Zee type contribution to the down-type quark chromo-EDM due to large $\tan \beta$ is attenuated.

\begin{figure}[htb]
\begin{center}
\includegraphics[width=8cm]{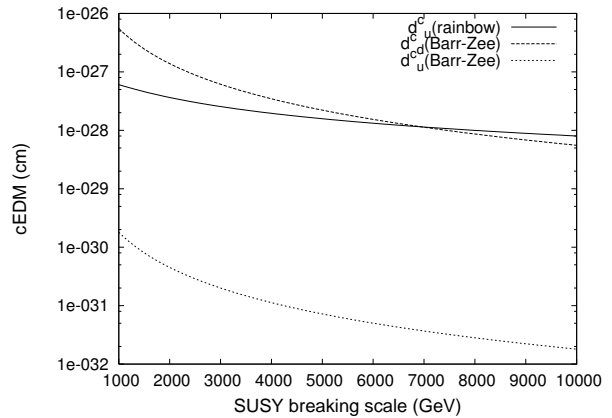}
\caption{\label{fig:rbcedm}
Chromo-EDM (absolute value) of the rainbow diagram plotted in the function of the supersymmetry breaking scale.
The Barr-Zee type contribution (absolute value) is also plotted for comparison.
}
\end{center}
\end{figure}

We have plotted in Fig. \ref{fig:rbcedm} the up quark chromo-EDM generated by the rainbow diagrams via the top-stop inner loop in the function of the supersymmetric scale $\Lambda_{\rm susy}$.
For comparison, we have also plotted the contribution of the Barr-Zee type diagram to the up and down quark chromo-EDM, assuming that both rainbow and Barr-Zee type contributions provide the maximal $CP$ violation, i.e., $\cos \theta_t \sin \theta_t \sin \delta_t \approx \cos \theta_t \sin \theta_t {\rm Im} ( e^{i\theta_\mu +i\delta_t} ) \approx \frac{1}{2}$.
We see that for sparticle masses near 1 TeV, the Barr-Zee type effect of the down quark is dominant.
Although being smaller, we can verify that the rainbow diagram is sizable compared with the Barr-Zee type diagram.
Moreover, the situation changes for heavier $\Lambda_{\rm susy}$.
The dominance of the rainbow type diagram in a heavier region can be explained by the dimensional analysis.
As we have fixed the $\mu$ parameter, the Barr-Zee type contribution scales as $\Lambda_{\rm susy}^{-2}$ which decreases faster than the rainbow diagrams which scale as $\Lambda_{\rm susy}^{-1}$ [see the gluino mass in the numerator of Eq. (\ref{eq:cedm})].
If the $\mu$ parameter has a supersymmetric dynamical origin \cite{giudice} and scales as $\Lambda_{\rm susy}$, the Barr-Zee type contribution dominates over the rainbow diagram over the whole region.
We should note also that the interference between the rainbow and Barr-Zee type diagrams is dependent on the relative phase and sign between the gluino mass parameter and the $\mu$ parameter.


To summarize the discussion, we have calculated the rainbow diagram contribution to the fermion EDM in the MSSM and analyzed the chromo-EDM contribution.
We have found that in many situations the rainbow diagrams provide a comparable contribution to the fermion chromo-EDM generated from the currently known Barr-Zee type process and, in some situations, can dominate over it.
The choice of the sign of the gaugino masses and the $\mu$ parameter can give rise to either destructive or constructive interferences.
If destructive interference occurs, significant suppression of $CP$ violation is also possible.
In this paper, we have not analyzed the EDM of leptons.
Actually, to complete the rainbow diagram contribution, we need also to consider the chargino propagation in the inner loop.
This will be the subject of our next work.


The author thanks H. Kamano and S. Ohkoda for useful help and advice.


\end{document}